\documentclass{iau} 
\usepackage{graphicx}

\title[The old massive open cluster NGC~6791] 
{Evidences of tidal distortion and mass loss from the old open cluster NGC~6791}

\author[Carraro \& Dalessandro]   
{Giovanni Carraro$^{1,2}$
 \and Emanuele Dalessandro$^3$}

\affiliation{$^1$ESO, Alondo de Cordova 3107, 19001, Santiago de Chile, Chile \\
$^2$Dept. of Physics and Astronomy, University of Padova, Vicolo Osservatorio 3, 35122, Padova, Italy\\email: {\tt gcarraro@eso.org} \\[\affilskip]
$^3$Dept. of Physics and Astronomy, University of Bologna, Via Ranzani 1, 40127, Bologna, Italy \\email: {\tt emanuele.dalessandr2@unibo.it}}

\pubyear{2015}
\volume{316}  
\jname{Formation, evolution, and survival of massive star clusters}
\editors{C. Charbonel and A. Nota, eds.}
\begin{document}

\maketitle

\begin{abstract}
We present the first evidence of clear signatures of tidal distortions in the density distribution of the fascinating open cluster NGC~6791.
We find that the 2D density map shows a clear elongation and an irregular distribution starting from $\sim 300^{\prime\prime}$  from the cluster center and two tails extending in opposite directions beyond the tidal radius. These features are aligned to both the absolute proper motion and to the Galactic centre directions. Accordingly we find that both the surface brightness and star count density profiles reveal a departure from a King model starting from $\sim600^{\prime\prime}$. These observational evidences suggest that NGC~6791 is currently undergoing  mass-loss likely due to gravitational shocking and interactions with the tidal field
of the Milky Way. We derive the expected mass-loss due to stellar evolution and tidal interactions and we estimate the initial cluster mass to be $M_{ini} = (1.5-4.0 ) \times 10^5 M_{\odot}$.

\end{abstract}

\firstsection 
\section{Context and results}
NGC~6791 is one of the most massive open cluster (Carraro 2014) in the MW, possibly harbouring more than a stellar population (Geisler et al. 2012, Bragaglia et al. 2014).We used deep images obtained with the wide field imager MegaCam mounted at the Canada-France-Hawaii Telescope (CFHT) to cover a $2^o \times 2^o$ area around the cluster (Dalessandro et al. 2015).  The upper left panel in Fig~1 shows the obtained ($g^{\prime}$, $g^{\prime}-r^{\prime})$ CMDs of the innermost region of NGC~6791 (left) and the external {\it control field} (right) including stars located at a distance $r \geq 3000^{\prime\prime}$ from the cluster center. Large scale 2D colour-coded surface density map of NGC6791 obtained by using the optical matched filter technique is shown in the upper right panel of Fig.~1.  The contour levels span from 3$\sigma$ to 40$\sigma$ with irregular steps. The solid arrow represents the direction of the absolute proper motion while the dashed ones mark the direction of the Galactic Center and that perpendicular to the  Galactic plane (Z=0).The observed star count density profile is shown in the upper left panel of Fig.~1 (open grey squares). The dashed line represents the density value of the background as obtained in the {\it control field}. The black filled dots are densities obtained after background subtraction. The best single-mass King model is also over-plotted to the observations (solid line). For $r \geq 600^{\prime\prime}$ the density profile clearly deviates from the King model following a power-law with exponent $\alpha \sim -1.7$.By using a simple analytic approach (Lamers et al. 2005), we estimated the mass likely lost by NGC~6791 during its evolution because of the effect of both stellar evolution and dynamical interactions. On this basis we estimated the cluster initial mass as a function of the dissolution time parameter $t_{0}$. In particular the red curve (lower right panel in Fig.~1) shows the dependence for the current cluster mass (5000 $M_{\odot}$) and its actual age of 8 Gyr. 

\begin{figure}[b]
\begin{center}
 \includegraphics[scale=0.25]{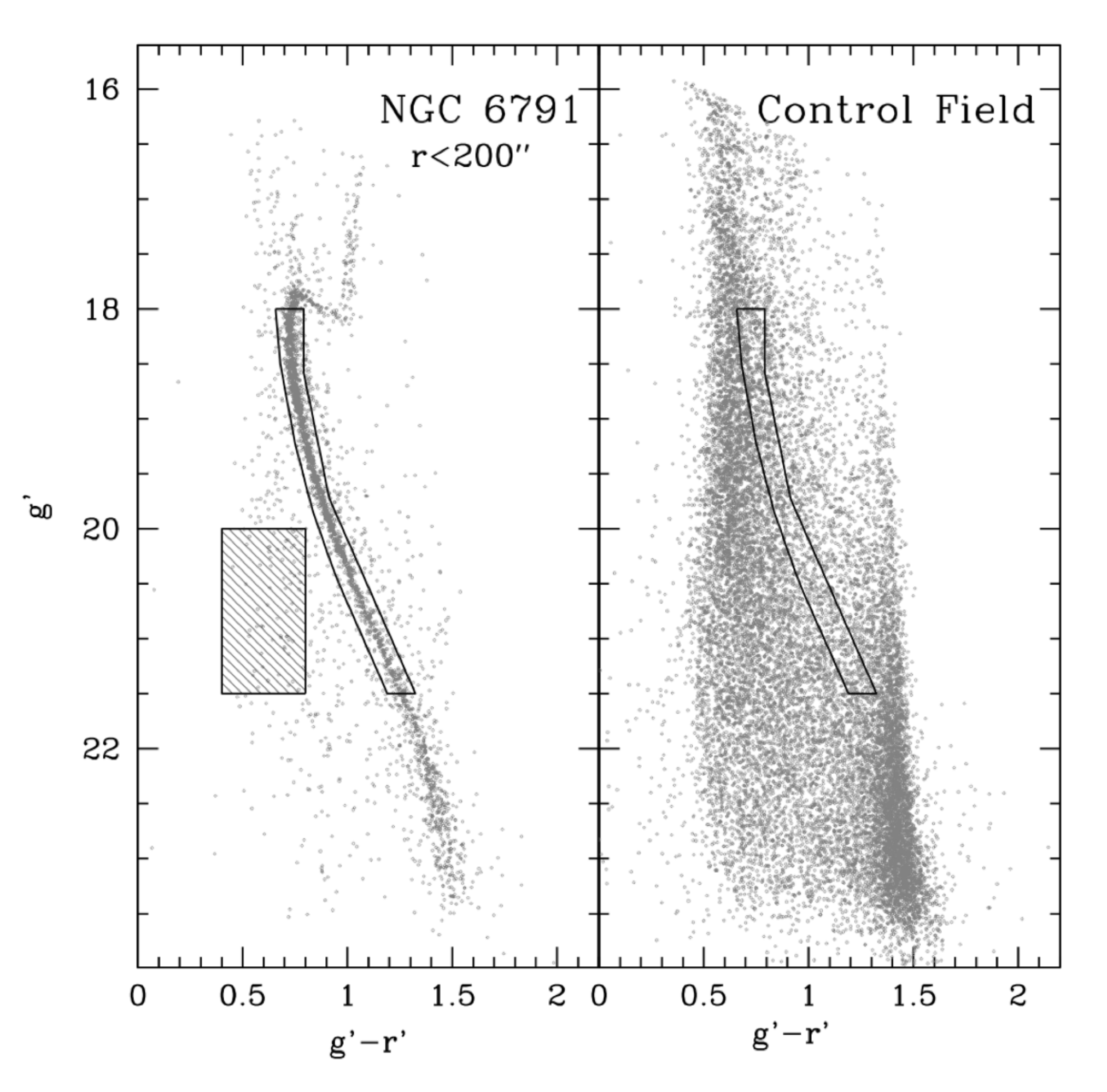} 
 \includegraphics[scale=0.25]{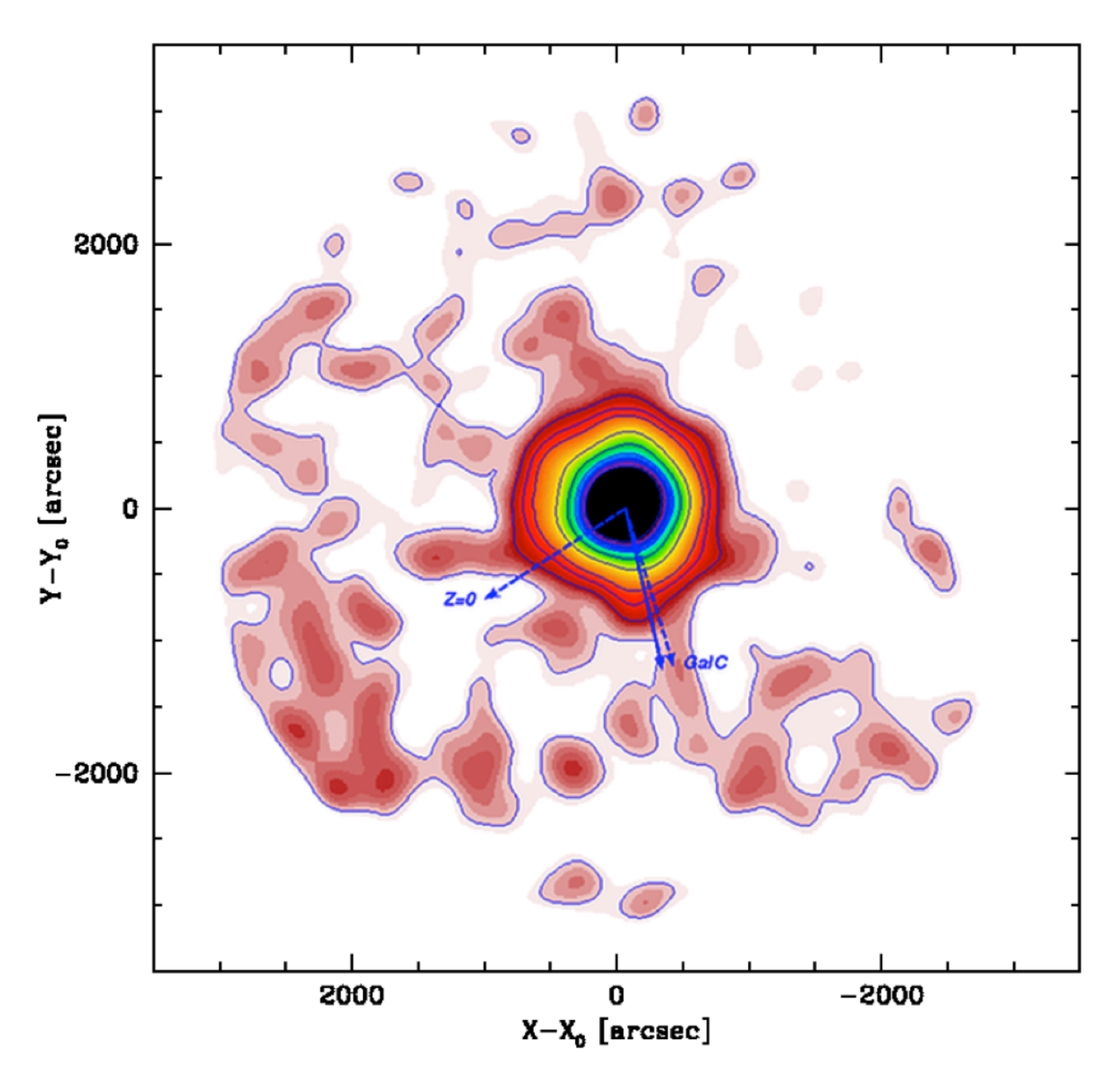} 
 \includegraphics[scale=0.25]{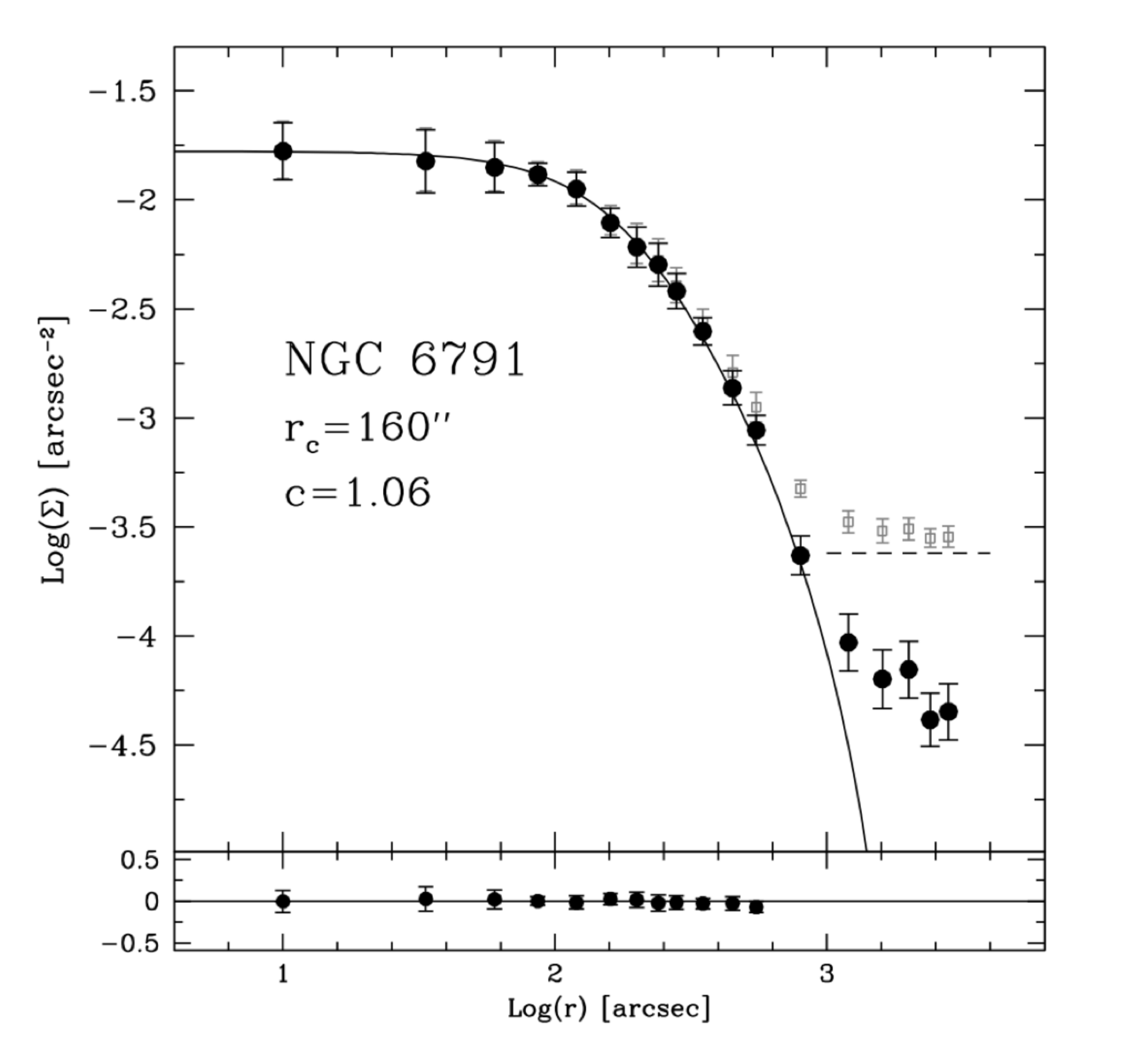} 
 \includegraphics[scale=0.25]{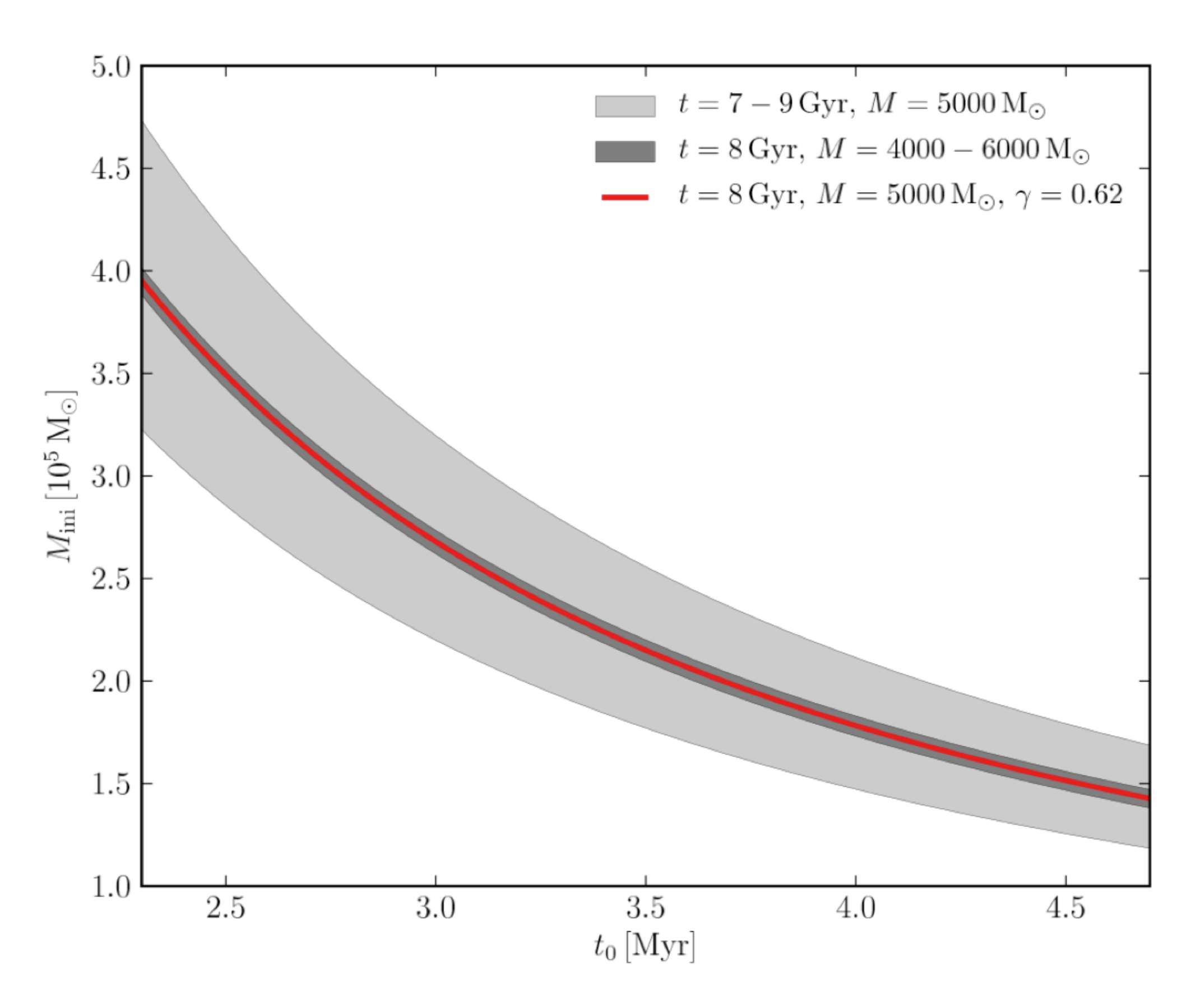} 
 \caption{{\bf Upper left}:  CMD of NGC~6791 and nearby field. {\bf Upper right}: star counts . {\bf Lower left}: King profile fitting . {\bf Lower right}: Mass at birth estimate. }
   \label{fig1}
\end{center}
\end{figure}

\section{Conclusions}
NGC~6791 shows clear evidence of tidal features in its star distribution in the form of irregular but evident elongations and tidal tails.
These features are present also in the star density and surface brightness profile and they represent clear indication of recent mass-loss.
By using the simple recipes we derived the initial mass of NGC6791 to be $M_{ini} = (1.5-4 ) \times 10^5 M_{\odot}$, i.e. several tens larger 
than its present day mass. This finding would qualitatively explain why the cluster could have survived for such a long time contrary to the expectations of  current estimates of  the destruction rate of Galactic open clusters.

\end{document}